\documentclass[acus]{JAC2000}

\usepackage{graphicx}

\begin{document}

\title{\flushright{TUAP066}\\[15pt] \centering 
IMPLEMENTATION OF THE EPICS DEVICE SUPPORT\\
FOR NETWORK-BASED CONTROLLERS
}

\author{K. Furukawa\thanks{e-Mail: \texttt{<kazuro.furukawa@kek.jp>}},
J. Chiba, N. Kamikubota, H. Nakagawa\\
High Energy Accelerator Research Organization (KEK),
Tsukuba, Ibaraki, 305-0801, Japan
}

\maketitle

\begin{abstract}

Network-based 
controllers such as PLC's and measurement stations will be used in 
the control system of the JAERI-KEK joint project (High Intensity 
Proton Accelerator Facility).  The ability for the network hardware 
and software to 
be standardized has led to this decision.  

EPICS software support for those controllers has been designed 
with special attention paid to robustness.  This software has been implemented 
and applied for the accelerator test stand, where the basic functionalities 
have been confirmed.  Miscellaneous functions, such as software diagnostics, 
will be added at a later date. 
To enable more manageable controllers, network-based equipment such as 
oscilloscopes are also being considered. 

\end{abstract}

\section{ INTRODUCTION }

Phase I of the JAERI-KEK joint project for high-intensity proton accelerators 
was recently approved for construction.  The detailed design of the control 
system is given elsewhere 
\cite{jkj-cont-ical2001}.  Because of the recent success of 
EPICS in the KEKB ring controls\cite{kekb-epics-pac99}
and the feasibility to share software resources of accelerator 
controls with other facilities, the EPICS control software 
environment\cite{epics} 
has been selected for use in the system.

Internet Protocol (IP) network controllers have been chosen for 
the controls of the linac portion of the project.  
The ability to use standard IP network software
and infrastructure for both controls and its management\cite{plc-we-ical99}
influenced our decision. 
If these controllers meet the performance requirements, as expected, 
their use may be extended to the entire project.  
  

This article describes the usage plan and the software implementation of 
such network-based controllers under EPICS. 

\section{ CONTROLLER USAGE UNDER EPICS }

A network-based controller and an EPICS 
IOC\footnote{IOC: input output controller.} may be connected with 
an IP network, as shown in Fig.~\ref{fig:net}.  In this example, a 
PLC\footnote{PLC: programmable logic controller.} is used, but
other network-based controllers act the same way.   
Five components in this scheme and their tasks are listed below: 

\begin{figure}[bt]
 \centering
 \includegraphics*[width=80mm]{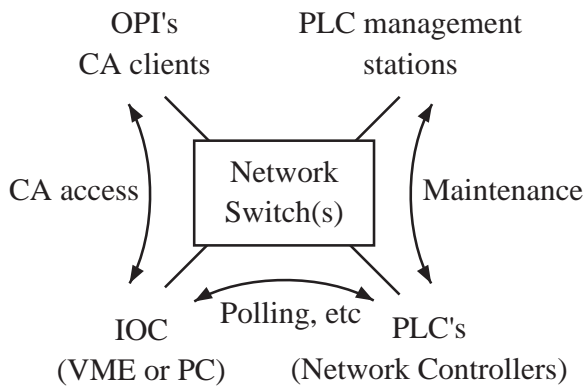}
 \caption{Same network shared between different purposes 
   between network controllers, EPICS IOCs, EPICS OPIs and management 
   stations.}
 \label{fig:net}
\end{figure}

\begin{Itemize}
\item The PLC controls local 
 equipment and carries local processing. 
 An equipment expert usually designs the PLC logic, and it may include
 a simple 
 local-operation panel.  It can be tested by a management station 
 without an EPICS environment.  This type of autonomous controllers is 
 useful when a robust control is required. 
\item 
 The IOC processes 
 logic between several PLCs, and keeps their current status in memory.  
 IOCs can be designed by an equipment expert or an operator. 
\item The OPI\footnote{OPI: operator interface.} has no knowledge of 
 network-based controllers, and therefore interacts with the IOC as 
 usual.
\item The management station is utilized to develop ladder software which 
 may be downloaded to a PLC, and to diagnose it.  The ability to test 
 from outside of the EPICS environment is useful in order to test for 
 errors in the EPICS database or applications themselves.  
\item Network hubs between them should be based on switch technology
 which does not suffer from message collisions.  The design of the 
 network topology is relatively flexible compared with other field 
 networks because of the IP network.  A connection to OPIs may 
 be isolated by a network router to limit the communication to PLCs 
 locally.  
\end{Itemize}

Figure~\ref{fig:net} is symmetric between them, since it shows a 
physical view.  Logically, PLCs are located on a local network while 
the OPIs are 
on a global network.  They communicate in three ways: 

\begin{Itemize}
\item Because a PLC cannot use the EPICS channel access (CA) protocol, 
 a PLC communicates with an IOC with its own protocol. 
 While this protocol is based on polling, an important 
 PLC can send urgent information to an IOC without being requested.  
\item The IOC communicates with OPIs through the normal CA protocol. 
\item The management station maintains PLCs during maintenance periods. 
 Due to the potential number of PLCs that will be in use, 
 it is important to manage them over the IP network. 
\end{Itemize}

\section{ NETWORK-BASED CONTROLLERS }

There are three types of network-based controllers being considered 
for the project:

\begin{Itemize}
\item Programmable logic controllers (PLC) for simple and medium-speed
 controls. 
\item Measurement stations (Yokogawa's WE7000) for medium-speed 
  waveform acquisition. 
\item Plug-in network controller boards for magnets with relatively large 
 power supplies. 
\end{Itemize}

VME modules installed in EPICS IOCs are used for other purposes, 
but new network equipment may be added.  Measurement equipment, such as 
network-based oscilloscopes, may be especially useful. 

\subsection{ PLCs }

At the electron linac in KEK, central control computers manage 
approximately 150 PLCs for rf, magnet, 
and vacuum controls.  
The PLC, FA-M3 (Factory ACE) from Yokogawa Co., was selected there
because of the network software reliability, and the ability to manage 
the PLCs over an IP network\cite{rf-cont-ical97,linac-subsys-ical2001}. 
Even the ladder 
software is able to be downloaded into a PLC over the network.  
We have decided to use the same 
type of PLCs at the joint project as well. 

The communication and control routines for PLCs were originally 
developed for use in a Unix environment.  While they were designed to access 
the shared-memory registers on the PLCs, they can also directly access 
I/O modules over the network. 

Since the routines were written with a generalized IP communication 
package\cite{s2}, they were  easily ported onto the VxWorks and Windows 
operating systems.  The routines on Windows machines are often useful 
for developers of ladder software of PLCs even without the EPICS environment.  

EPICS device support software has been  written utilizing those routines 
on VxWorks.  It provides standard EPICS access methods which can be 
called from any channel access (CA) clients.  Registers are read by and 
written to the PLCs.  Each of registers is specified by an INP/OUT 
EPICS record field, which uses an IP address, or a host name, and a 
register address. 

\begin{figure}[bt]
 \centering
 \includegraphics*[width=80mm]{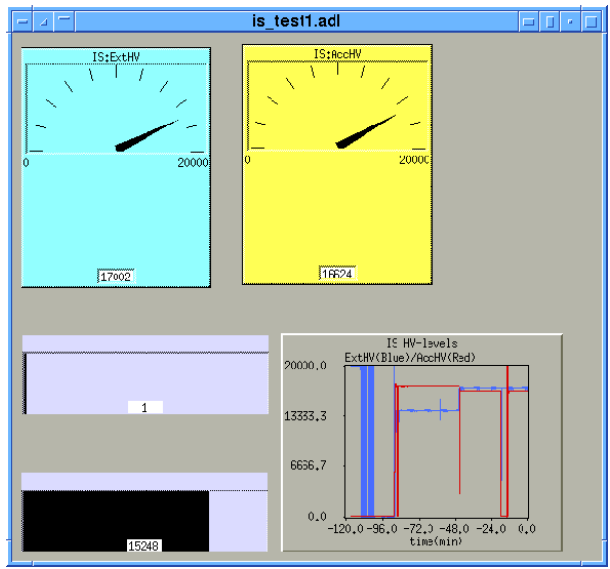}
 \caption{Example panel for testing the ion source of the linac, 
   which utilizes the EPICS PLC records.}
 \label{fig:medm}
\end{figure}

The MEDM panel shown in Fig.~\ref{fig:medm} provides an example of a 
channel access client.  The panel displays 
current high voltage values and their strip charts for the ion source 
being conditioned in the linac. 

This type of application can be handled by the current software version 
without 
problems, but the 
current implementation of the device support software is not 
yet optimal. A plan to upgrade the software with a conditional write 
function, which is described 
later, may be necessary.  

\subsection{ Measurement Stations }

A waveform acquisition is essential to beam instrumentation 
and microwave measurements.  The measurement station, 
WE7000 from Yokogawa Co.\, has been well adopted to beam instrumentation 
at KEK\cite{we7000-ical97}.  
In addition, their cost performance and electro-magnetic noise elimination 
are promising.    

Three types of waveform digitizers (100ks/s, 100Ms/s and 
1Gs/s) are currently considered.  We decided to out-source the EPICS 
device support software, because we thought that it would be a good 
example of out-sourcing.  Although it took some time for the company to 
understand the EPICS software environment, waveform records have 
successfully built 
using disclosed information from Yokogawa.  
At this time , we are evaluating the 
performance of the software. 

\subsection{ Plug-in Network Controller }

While designing the magnet power supplies for the drift-tube linac (DTL) 
and the separated-DTL, it was realized that a special type of 
controller was needed, since power supplies were intelligent and had many 
functions. 

we thus designed a plug-in-type network controller board, which 
transfers information and commands over the IP network and a  
local processor located inside a power supply.  There are approximately 
50 registers, half of which are utilized for network communication 
and for diagnostic purposes, such as the last IP 
address accessed. 

The controller boards are being built with the power supplies and 
will be evaluated soon.  The software will be very similar and compatible with 
the PLCs'. 

\section{ CONSIDERATION }

Since network-based controllers may reside on a global network, we 
should be very careful about programming and configuring them. 
Although the number of persons who access such controllers were 
limited in the previous project, this may not be the case in the present 
new project. 
Therefore, we decided to make several rules for use of the controllers:

\begin{Itemize}
\item We will put an unique identification number (ID) to 
each PLC and plug-in network controller.  Since it will be written in 
hardware or ladder software, a mistake in the configuration of the 
IP-address may be found from a management station.
\item A clock counter of the controller should be consulted routinely 
from a management station to ensure that it works properly. 
\item While read functions are not restricted, write functions should 
be limited to some range of register addresses.  For important controllers, 
a value should be always written indirectly with a value and an address. 
\end{Itemize}


\section{ CONCLUSION }

The combination of the network-based controllers with the EPICS toolkits 
will 
enhance the manageability of the control system.  The software for 
EPICS toolkits has been developed and is currently being tested. 
They will soon be used in commissioning the first part of the linac. 

\section{ ACKNOWLEDGMENTS }

The authors would like to thank the KEKB ring control group people 
and the joint project staff for valuable discussions.

\end{document}